\documentclass[pra,aps
,preprint
,tightenlines
 ]{revtex4}

\usepackage{graphicx}

\begin{document}
	 
 \def\bra#1{\mathinner{\langle{#1}|}}
\def\ket#1{\mathinner{|{#1}\rangle}}
\def\braket#1{\mathinner{\langle{#1}\rangle}}
\def\Bra#1{\left<#1\right|}
\def\Ket#1{\left|#1\right>}
{\catcode`\|=\active 
  \gdef\Braket#1{\left<\mathcode`\|"8000\let|\BraVert {#1}\right>}}
\def\BraVert{\egroup\,\mid@vertical\,\bgroup}
% The \mid@vertical is \vrule with ordinary TeX but \middle| in eTeX.
% We always avoid a \mathchoice in making the inner vertical lines.  
% Note that \right>, prints the same as \right\rangle but is faster.  
%
% \def\ketbra#1#2{\ket{#1}\bra{#2}}
% \def\Ketbra#1#2{\left|{#1}\vphantom{#2}\right>\left<{#2}\vphantom{#1}\right|}

% \Set{...|...} Only the first | is treated specially.
{\catcode`\|=\active
  \gdef\set#1{\mathinner{\lbrace\,{\mathcode`\|"8000\let|\midvert #1}\,\rbrace}}
  \gdef\Set#1{\left\{\:{\mathcode`\|"8000\let|\SetVert #1}\:\right\}}}
\def\midvert{\egroup\mid\bgroup}
\def\SetVert{\egroup\;\mid@vertical\;\bgroup}

% If the user is using e-TeX with its \middle primitive, use that for
% verticals instead of \vrule.
%
\begingroup
 \edef\@tempa{\meaning\middle}
 \edef\@tempb{\string\middle}
\expandafter \endgroup \ifx\@tempa\@tempb
 \def\mid@vertical{\middle|}
\else
 \let\mid@vertical\vrule
\fi

\title{Comparison of LOQC C-sign gates with ancilla inefficiency and an
improvement to functionality under these conditions}

\author{A. P. Lund}
\author{T. B. Bell}
\author{T. C. Ralph}
\affiliation{Centre for Quantum Computer Technology, Department of Physics
\\ University of Queensland, QLD 4072, Australia
\\ Fax: +61 7 3365 1242  Telephone: +61 7 3365 3412 \\
email: lund@physics.uq.edu.au\\}

\begin{abstract}
We compare three proposals for non-deterministic C-sign gates implemented 
using linear
optics and conditional measurements with non-ideal ancilla mode 
production and detection.  The simplified
KLM gate~[Ralph et al, Phys.Rev.A {\bf 65}, 012314 (2001)] 
appears to be the most resilient under these conditions.
We also
find that the operation of this gate can be improved 
by adjusting the beamsplitter ratios to
compensate to some extent for the effects of the imperfect ancilla.
\end{abstract}

\maketitle

\section{Introduction}

Linear optics Quantum Computation (LOQC) \cite{kni00} offers an
elegant way of implementing quantum gates on optical qubits using the
inherent non-linearity of conditional measurements. This is achieved
by introducing ancilla photons which interact with the linear circuit
and are then detected. However, it has been shown that the accuracy of
the gate operation is strongly dependent on the quality of the
detectors used to detect the ancilla photons \cite{glancy}.

Three distinct architectures have now been suggested for implementing the
fundamental two qubit gate, the control-sign (C-sign) \cite{ral00,kni01,pit00}.
It is natural to ask firstly whether all these architectures are
equally sensitive to ancilla detector efficiency and secondly if it
is possible to optimize gate operation to counter (to some extent) the
effects of detector inefficiency. In this paper we address these
questions and include in our analysis the converse issue of
inefficiency in ancilla production.

We begin in section~\ref{analysis} by presenting our analysis technique. 
In section~\ref{comparisons} we introduce the three versions of the
C-sign gate and then present our comparative analysis. In
section~\ref{tweek} we discuss improvements to the least 
sensitive of these gates.  We conclude in section~\ref{conc}.

\section{\label{analysis}Gate Analysis}

In performing the analysis of the gates we consider ideal qubits sent 
into a non-deterministic LOQC gate consisting of a linear optical 
circuit interacting with prepared ancilla modes.  The ancilla modes 
are then detected and the state at the output modes is kept if the 
measurement successfully matches the condition required for correct 
operation.  It is assumed that mode matching errors and loss in the 
optical circuit can be neglected, but that inefficiency in the 
production and detection of the ancilla cannot be neglected.  When the 
ancilla detection result indicates successful gate operation
the output state is compared with the expected output via their 
fidelity
\[
\braket{\Psi_{exp}|\rho_{out}|\Psi_{exp}}
\]
where $\rho_{out}$ is the output density operator and $\ket{\Psi_{exp}}$ is 
the expected output.  The fidelity is calculated in this way for all 
input states and the minimum fidelity is found.  This is then taken as 
the figure of merit used for comparison.  Under ideal conditions the 
fidelity is one for all inputs 
but lower numbers indicate reduced accuracy of the 
gate.  Inefficient production and detection in ancilla modes are 
expected to have two effects: reduction in the probability of 
successful gate operation and a reduction in the fidelity when 
successful operation occurs.

Detector and input inefficiencies are simulated by introducing a 
beamsplitter with a reflectivity equal to the efficiency.  The 
refected mode of each beamsplitter remains in the system and the 
transmitted mode is lost.  No information can be retrieved in the loss 
mode so a partial trace is performed over this mode leaving the system 
in a mixed state.  
% This situation is shown in 
% figure~\ref{detector}. 
% %
% \begin{figure}
%   \includegraphics[width=3.5cm]{Detector.eps}
%   \caption{\label{detector} The simulation of a non-ideal
%   detector or non-ideal input.
%   The beamsplitter ratio is equal to the efficiency and unused mode is
%   traced over. }
% \end{figure}

\begin{figure*}[hbtp]
  \includegraphics[width=0.7\textwidth]{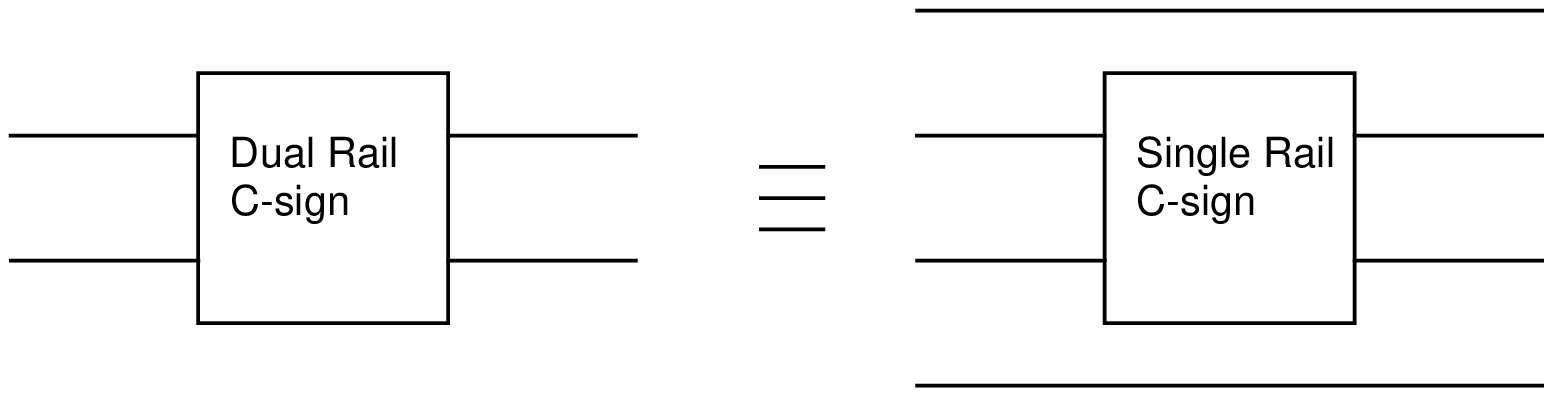} 
  \caption{\label{dual-single}This diagram shows that the dual rail 
  form is equivalent to the single rail form with extra modes.  In 
  dual rail format the lines represent two modes usually the 
  horizontal and vertical modes of a single spatial mode.  In the 
  single rail format the lines represent one mode and the qubit is 
  encoded in the photon number.}
\end{figure*}

For the sake of computational simplicity all the gates are analyzed in 
a single rail format \cite{lun00}, where the zero photon state 
$\ket{0}$ represents logical zero and the single photon state 
$\ket{1}$ represents logical one.  In single rail format the C-sign 
operation is defined by: 
\begin{eqnarray*}
\ket{0}\ket{0} & \longrightarrow & \ket{0}\ket{0} \\
\ket{0}\ket{1} & \longrightarrow & \ket{0}\ket{1} \\
\ket{1}\ket{0} & \longrightarrow & \ket{1}\ket{0} \\
\ket{1}\ket{1} & \longrightarrow & -\ket{1}\ket{1}
\end{eqnarray*}
Single qubit manipulations are difficult using single rail logic.  
Thus dual rail logic \cite{mil88} is normally adopted in practice with 
the qubit defined across two optical modes.  The logical zero is 
represented by a single photon occupation of one mode with the other 
in the vacuum state.  The logical one is the reverse of the logical 
zero state with a single photon in the other mode.  In LOQC, dual rail 
logic is often implemented using the horizontal and vertical 
polarization modes of a single spatial mode.  For the special case of 
a C-sign gate the dual rail form is equivalent to the single rail 
form, just with added modes which do not participate in any 
interactions (see figure~\ref{dual-single}).
This can be seen from the definition of C-sign operation in the dual 
rail format (written in photon occupation form): 
\begin{eqnarray*}
(\ket{1}\ket{0})(\ket{1}\ket{0}) & \longrightarrow &
(\ket{1}\ket{0})(\ket{1}\ket{0}) \\
(\ket{1}\ket{0})(\ket{0}\ket{1}) & \longrightarrow &
(\ket{1}\ket{0})(\ket{0}\ket{1}) \\
(\ket{0}\ket{1})(\ket{1}\ket{0}) & \longrightarrow &
(\ket{0}\ket{1})(\ket{1}\ket{0}) \\
(\ket{0}\ket{1})(\ket{0}\ket{1}) & \longrightarrow &
-(\ket{0}\ket{1})(\ket{0}\ket{1}).
\end{eqnarray*}
The first two bracketed states represent the first qubit while the 
second two represent the second qubit.  Note that if the first mode is removed 
from all the qubits in the dual rail format then the single rail 
format is obtained.  Because the extra modes do not participate in 
C-Sign gates (the assumed sources of loss are not present), single 
rail and dual rail fidelities are identical.  Once in the dual rail 
format Control-NOT operation can be constructed by mixing the two 
target modes (the modes on which the controlled operation is to be 
applied) on a 50:50 beamsplitter before and after the C-sign 
operation.

The fidelity of each of the gates was calculated as follows.  The 
operator evolution equations of each particular gate were calculated 
and inverted.  
The density operator for the required input state (including ancilla) 
was evolved using the solutions from the inverted equations.  The 
loss modes are traced over, and detected modes are projected 
onto the required state.  The remaining density operator 
$\hat{\rho}_{out}$ 
describes the output state which is now normalized to have 
$\textrm{Tr}(\hat{\rho}_{out}) = 1$.  This renormalization is because we 
only wish to consider the accuracy of the gate assuming a successful 
detection event; the success rate is considered separately.  The 
fidelity of the gate is calculated by finding the minimum of 
$\bra{\Psi_{exp}}\hat{\rho}_{out}\ket{\Psi_{exp}}$ over all input states 
where $\ket{\Psi_{exp}}$ is the expected output state from the used 
input state $\ket{\Psi_{in}}$.  The general input state 
$\ket{\Psi_{in}}$ was written as follows:
\begin{eqnarray}
\ket{\Psi_{in}§} & = & \cos{\alpha} \ket{00} + \sin{\alpha} \cos{\beta} 
\ket{10} + \\* \nonumber && \sin{\alpha} \sin{\beta} \cos{\gamma} 
\ket{01} + \sin{\alpha} \sin{\beta} \sin{\gamma} \ket{11} .
\end{eqnarray}
The advantage of writing the state in this form is that the optimization 
for finding the minimum fidelity can be performed over the variables 
$\alpha$, $\beta$ and $\gamma$ instead of a constrained optimization.

\section{\label{comparisons}Gate Comparisons}

The three C-sign gates that were compared in our analysis are as 
follows:
\begin{description}
\item[KLM] The original non-deterministic C-sign gate introduced by 
Knill, Laflamme and Milburn~\cite{kni00} is based on the operation of 
the so-called non-linear sign shift (NS) gate, which performs the 
transformation 
\begin{equation}
\label{NSeqn}
\alpha\ket{0} + \beta\ket{1} + \gamma\ket{2} \longrightarrow \alpha\ket{0} +
\beta\ket{1} - \gamma\ket{2} .
\end{equation}
A simplification of the original design, shown in figure~\ref{NSgate}, 
was introduced by Ralph \emph{et.  al.}~\cite{ral00} and is used in 
our calculations.  Vacuum ($0$) and single ($1$) photon states are 
injected into the ancilla modes.  The gate succeeds when the output 
ancilla are detected to be in the same state as was injected.  C-sign 
operation is achieved by placing an NS gate in each arm of a balanced 
Mach Zehnder interferometer as shown in figure~\ref{csklm}. 
\begin{figure*}
  \includegraphics[width=7cm]{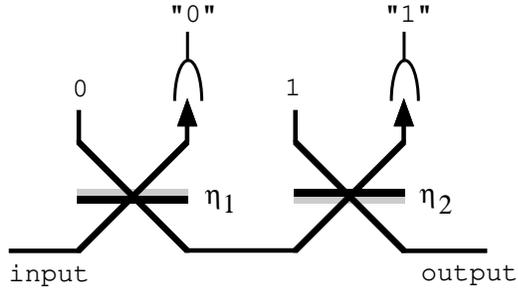} 
  \caption{\label{NSgate}The non-deterministic gate which performs the 
  operation described by equation~\ref{NSeqn}.  The beamsplitter 
  reflectivities are $\eta_{1}=5 - 3 \sqrt{2}$ and $\eta_{2} = (3 - 
  \sqrt{2})/7$.}
\end{figure*}
Photon bunching in the interferometer then produces the sign shift 
when both control and target modes are in the $\ket{1}$ state.  The 
probability of success for the gate is approximately $1/20$.

\begin{figure*}
  \includegraphics[width=11cm]{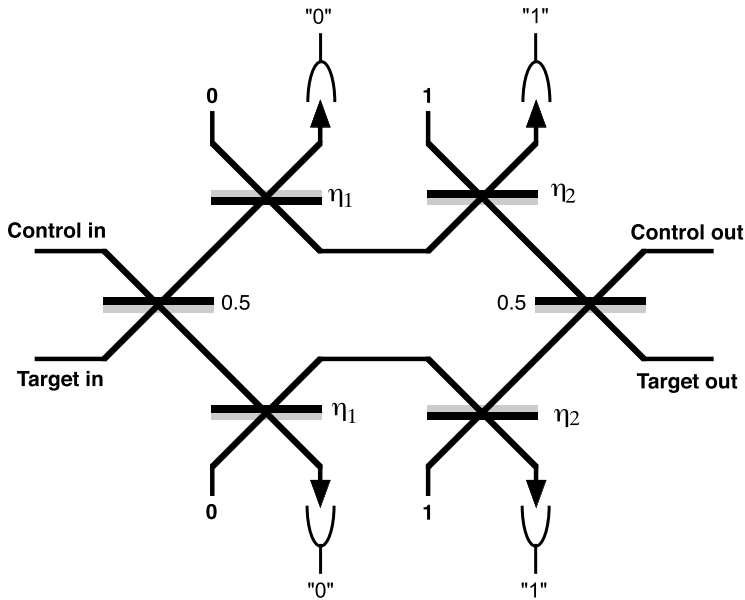}
  \caption{\label{csklm} The (simplified) KLM control sign gate
  \cite{ral00}. The
  unnumbered beamsplitters have reflectivities $\eta_{1}=5 - 3
  \sqrt{2}$ and $\eta_{2} = (3 - \sqrt{2})/7$.}
\end{figure*}

\item[Knill] Our second gate shown in figure~\ref{csknill} was 
introduced by Knill~\cite{kni01}.  It directly implements the C-sign 
operation.  In contrast to the KLM gate it has no classical 
interferometric elements and requires only two ancilla, both prepared 
in single photon states.  The gate succeeds when the output ancilla 
are both measured to be single photon states.  The probability for 
success of the Knill gate is $1/13.5$.

\begin{figure*}
  \includegraphics[width=9cm]{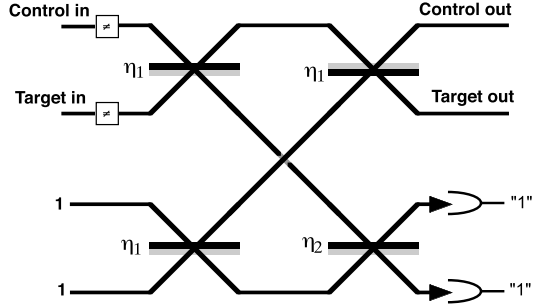}
  \caption{\label{csknill} The knill control sign gate \cite{kni01}.
  The reflectivities are $\eta_{1} = \frac{1}{3}$ and $\eta_{2} =
  \frac{1}{6}(3+\sqrt{6})$.  Note the beamsplitter convention here is 
  different (see text).}
\end{figure*}

\item[PJF] Our third gate was introduced by Pittman, Jacobs and 
Franson ~\cite{pit00} and is shown in figure~\ref{cnotpittman}.  A 
related gate is that introduced by Koashi \emph{et.  
al.}~\cite{koashi}.  Unlike the other two gates, the PJF gate 
requires entanglement between the two ancilla modes.  All 
beamsplitters have a reflectivity of 0.5.  For the ancilla modes which 
are detected, the pairs of detectors shown must have exactly one 
photon, total, in the two modes for the gate to succeed.  Rotations to 
the output may be necessary depending on which mode the single photon 
is found.  The gate functionality is driven by the entanglement in the 
ancilla modes.  The state of the four ancilla modes is (in the form 
$\ket{a_{1} a_{2} a_{3} a_{4}}$) $(1/\sqrt{2}) (\ket{0110} + 
\ket{1001})$.  The probability of success of the PJF gate is 
$1/4$.

\begin{figure*}
  \includegraphics[width=10cm]{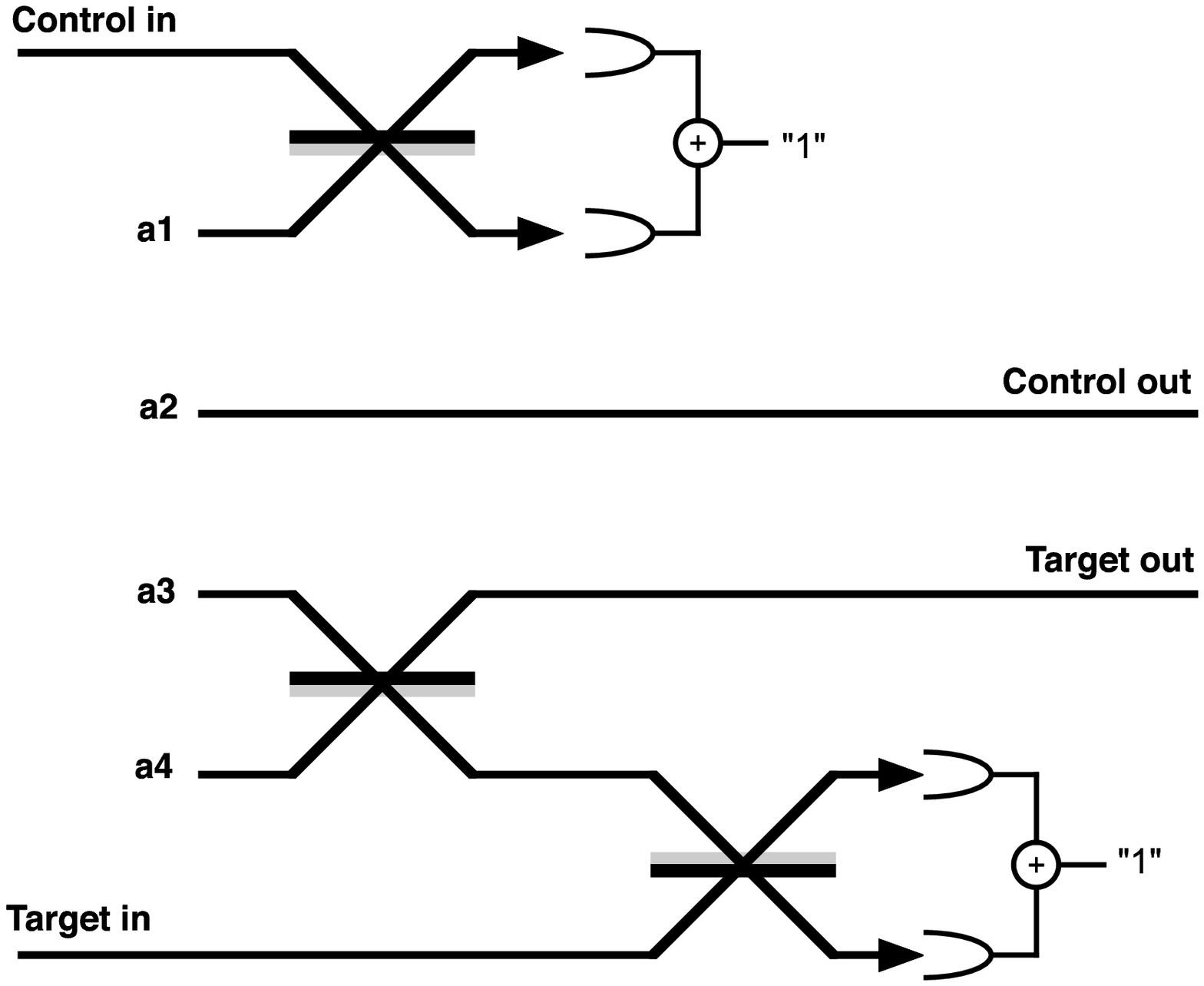}
  \caption{\label{cnotpittman} The Pittman, et. al. C-sign gate
  \cite{pit00}.  The schematic here uses normal beamsplitters with
  reflectivities of $0.5$.  The detector pairs must measure one photon
  in total. The ancilla modes are prepared as $\ket{a_{1} a_{2} a_{3}
  a_{4}} \rightarrow (1/\sqrt{2}) (\ket{0110} + \ket{1001})$}
\end{figure*}

\end{description}
Throughout the remainder of this paper, the gates will be called by
the names just introduced. Note that the beamsplitter
conventions differ between the proposals.
The KLM and PJF gates
have beamsplitters which have a sign change on reflection off the grey
side but the Knill gate has a sign change on transmission for beams
incident on the black side.

\begin{figure}
  \includegraphics[width=0.4\textwidth]{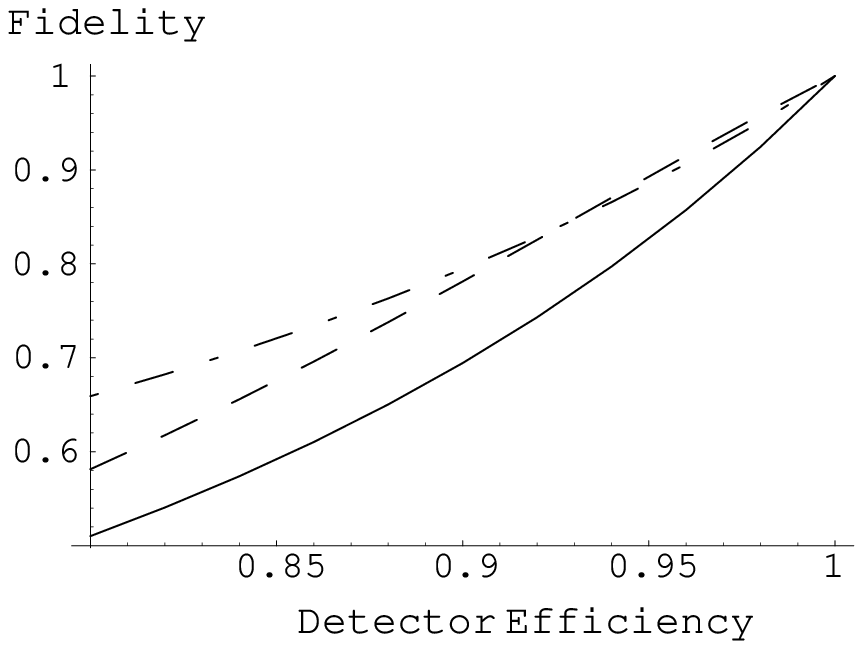}
  \caption{\label{det-comp} A comparison of the minimum fidelity of the three
  gates (PJF - solid, Knill - dashed, KLM - dot-dashed) as a
  function of the detector efficiencies.}
\end{figure}

Figure~\ref{det-comp} shows the results of the fidelity calculations 
(as described in the previous section) for the three gates when only 
the detectors exhibit loss (i.e.  perfect state input).  The parameter 
along the abscissa is the detector efficiency and the ordinate shows 
the fidelity of the gate at that efficiency.  The solid line 
represents the PJF gate, the dashed line shows the Knill 
gate and the dot-dashed line shows the KLM gate.

All gates show a quite steep decrease in minimum fidelity as a 
function of efficiency, illustrating the sensitivity of LOQC gates to 
this sort of loss (recall though that this is minimum fidelity and so 
represents a worst case scenario). For detector efficiencies greater than about 93\% the Knill gate gives
marginally better performance, but for detector efficiencies below this
value the KLM gate shows a better fidelity by a significant margin.

\begin{figure}
  \includegraphics[width=0.4\textwidth]{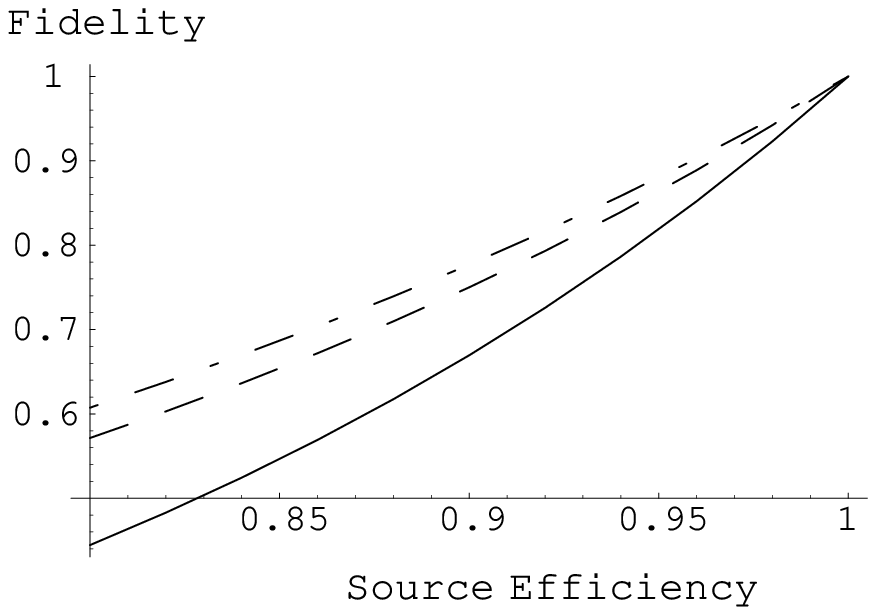}
  \caption{\label{input-comp} A comparison of the minimum fidelity of the three
  gates (PJF - solid, Knill - dashed, KLM - dot-dashed) as a
  function of the input efficiencies.}
\end{figure}

A similar analysis can be done with ancilla production efficiencies.
Figure~\ref{input-comp} shows this analysis and has the same
gate - plot~style correspondence as in figure~\ref{det-comp}. Once 
again a steep decrease in minimum fidelity as a function of 
efficiency is observed.
In this figure it can be seen that the KLM gate has the highest 
minimum fidelity for the range of efficiencies shown.  From the 
figures we 
may conclude that, as assessed by minimum fidelity, 
the simplified KLM gate is in general the most forgiving in 
the presence of ancilla production and detection inefficiencies.

\section{\label{tweek}Gate Fidelity Improvement}

One effect of reduced ancilla efficiency is to bias the 
probability of successful gate operation for different input states.  
This is a detrimental effect as some information about the input state 
is thus leaked through the statistics of the projective measurements 
success. In turn this results in biasing of the fidelities of the gate 
for different inputs. For example with the KLM gate the fidelity for 
the
$\ket{0}\ket{0}$ input state is unaffected by ancilla inefficiencies 
while the $\ket{0}\ket{1}$ and $\ket{1}\ket{0}$ states are most strongly 
affected, with these states giving the minimum fidelity for this gate. 
This suggests it may be possible to improve upon the 
fidelity gained here if one were to adjust the elements in the gate to 
compensate for the biasing of gate functionality 
incurred due to the ancilla inefficiencies.  Using this idea as a guide 
we have improved the performance of the KLM gate.

The KLM gate is constructed from two NS gates, which ideally 
perform the operation given in equation~\ref{NSeqn}.  The gate has two 
parameters which can be altered:  the reflectivities of each of 
the two beamsplitters.  Using the same technique as above 
for calculating the gate fidelity, we can optimize the 
fidelity with respect to these beamsplitter ratios for fixed detector 
and input efficiency.  It is assumed that the two NS gates in the 
whole C-sign gate have the same beamsplitter ratios, maintaining the 
symmetry of the gate.

\begin{figure*}
  \includegraphics[width=11cm]{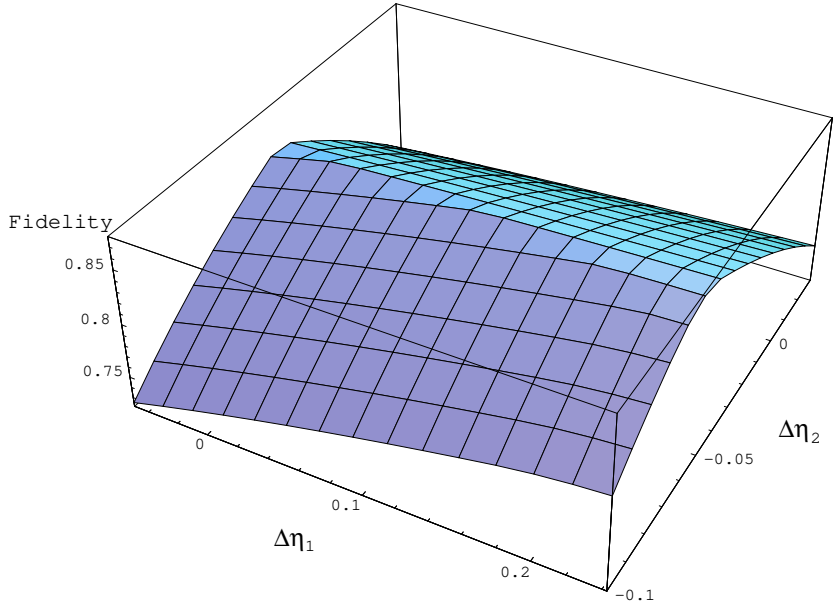}
  \caption{\label{3Ddet}This plot shows the minimum 
  fidelity (z-axis) of the modified
  KLM gate with detector efficiency of 0.9 and perfect ancilla input
  for a range of beamsplitter ratios of the two NS gates (x-y axes).
  The x-y axes show the change in the reflectivity from the normal
  reflectivities of the NS gate.
  The important feature of this plot is the increasing fidelity as $\eta_{1}$
  is increased.  The most positive value of $\eta_{1}$ shown here is
  the maximum value it can take.}
\end{figure*}

Figure~\ref{3Ddet} shows the fidelity of the simplified KLM gate when 
the prepared ancilla are kept the same and the detection scheme is the 
same as proposed, but the beamsplitter ratios in the two NS gates are 
varied.  The `$\Delta \eta 1$' and `$\Delta \eta 2$' axes show the 
change in the beamsplitter ratios from their initial values; that is, 
for the point (0,0) the beamsplitter ratios have not changed.  The 
z-axis shows the fidelity of the gate.  The assumed loss with this 
diagram is 90\%
detector efficiency and perfect input efficiency.

The important feature of this plot is the increase of fidelity with 
$\eta_{1}$.  To the far right of the $\eta_{1}$ axis is the limit of 
the allowed values for $\eta_{1}$.  This limit is imposed by the 
necessity that reflectivities lie between zero and one.  So in this 
case, the fidelity can be optimized by choosing the first beamsplitter 
perfectly reflective.  Doing this, in effect, removes the detector 
which measures zero photons and removes the vacuum input.  Inefficient 
equipment is removed from the gate and the gate complexity is reduced.  
All that remains is to optimize the fidelity along the `$\Delta \eta 
2$' axis.  This feature of increasing fidelity with $\eta_{1}$ is seen 
here with detector efficiencies up to about 99\%.

The increasing fidelity with $\eta_{1}$ is not seen with a lossy 
source.  However, when the source efficiency drops slightly below 
unity the relationship between the gate fidelity and $\eta_{1}$ is 
almost flat.  For source efficiency of about 98\%, the improvement in 
the fidelity is only about $0.01$ at the actual optimized value of 
$\eta_{1}$ and $\eta_{2}$ compared with setting $\eta_{1}=1$.  So for 
simplicity, the fidelity will be considered optimized at $\eta_{1}=1$ 
for both lossy sources and detectors.

\begin{figure*}
  \includegraphics[width=0.5\textwidth]{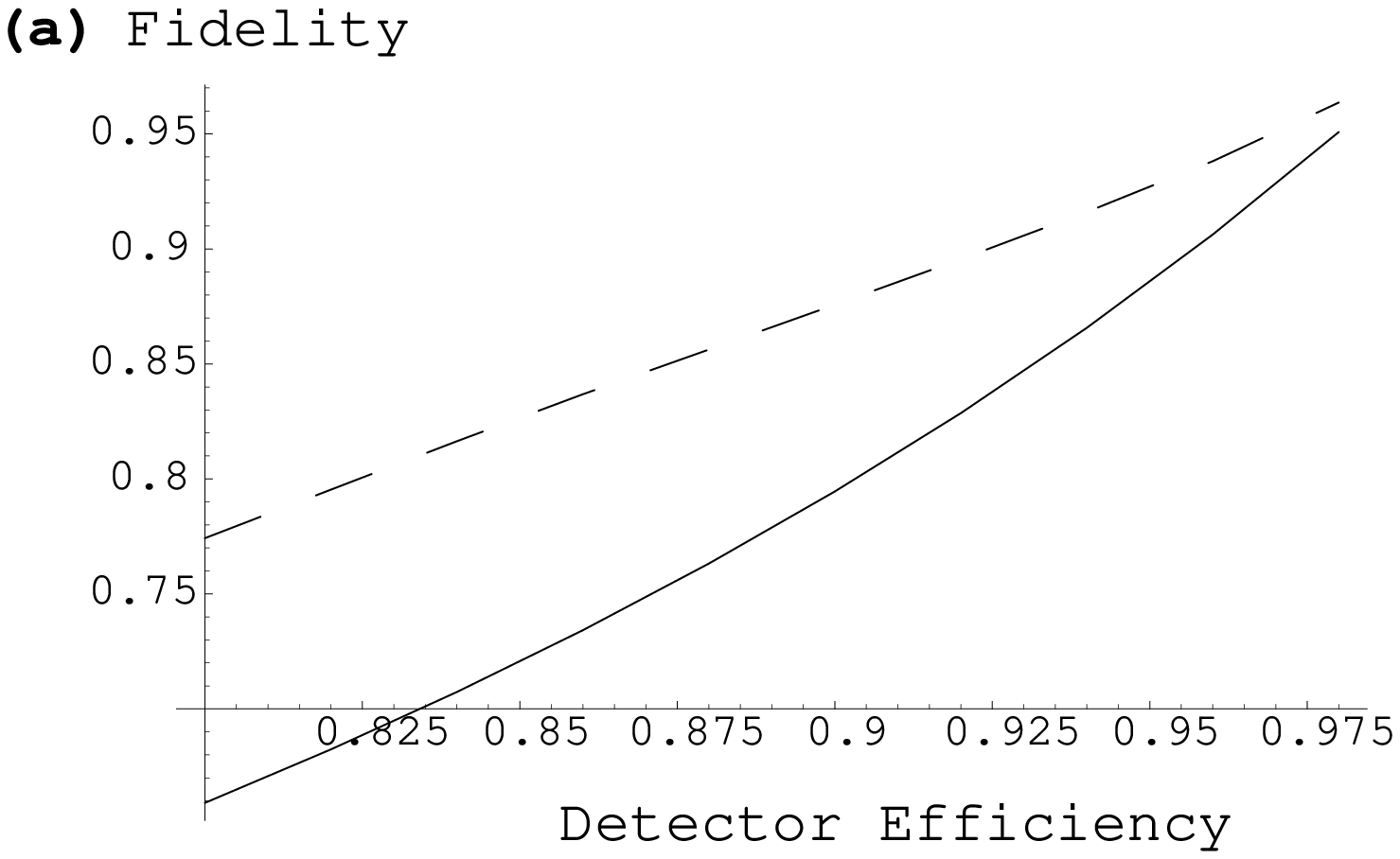}
  \includegraphics[width=0.4\textwidth]{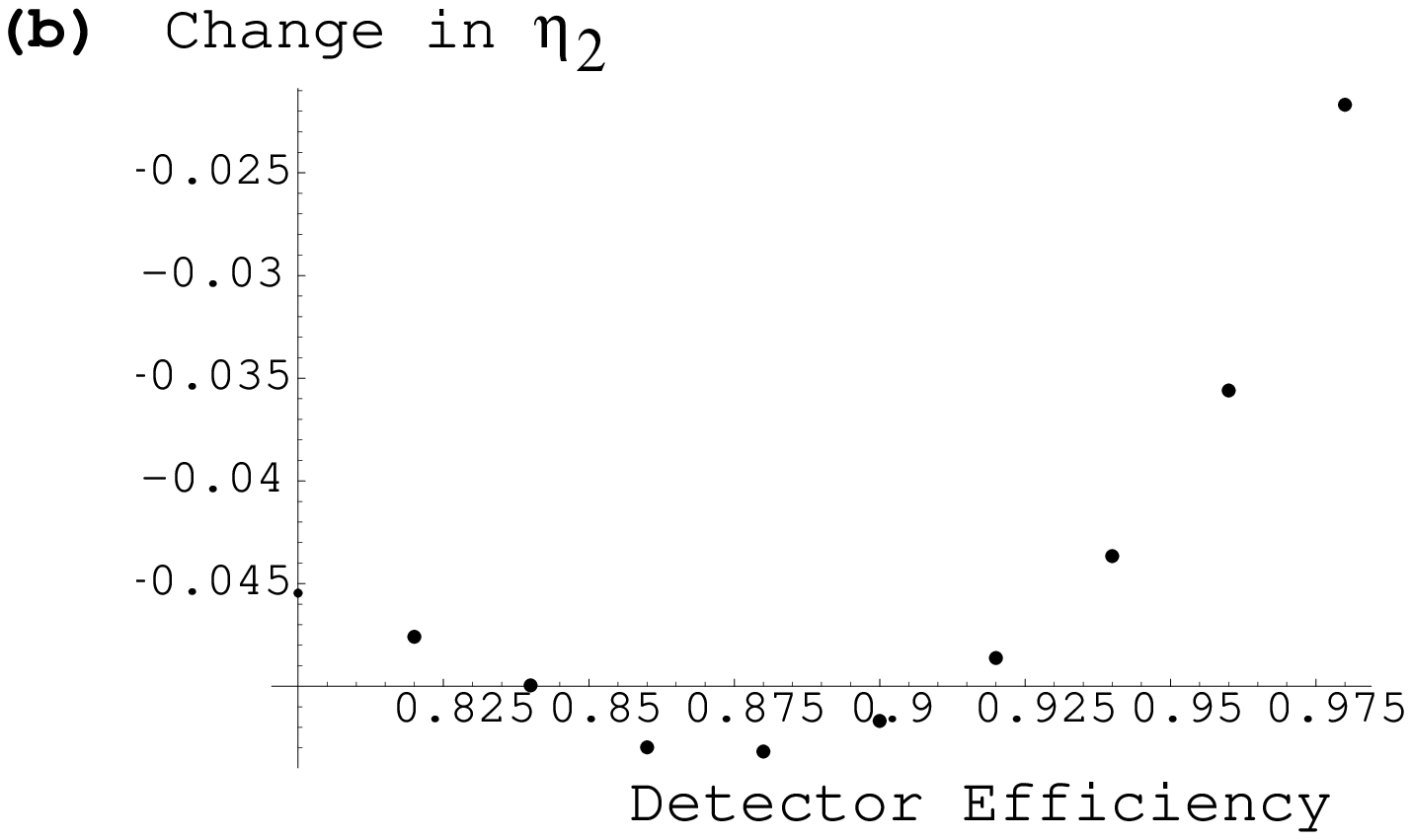}

  \includegraphics[width=0.4\textwidth]{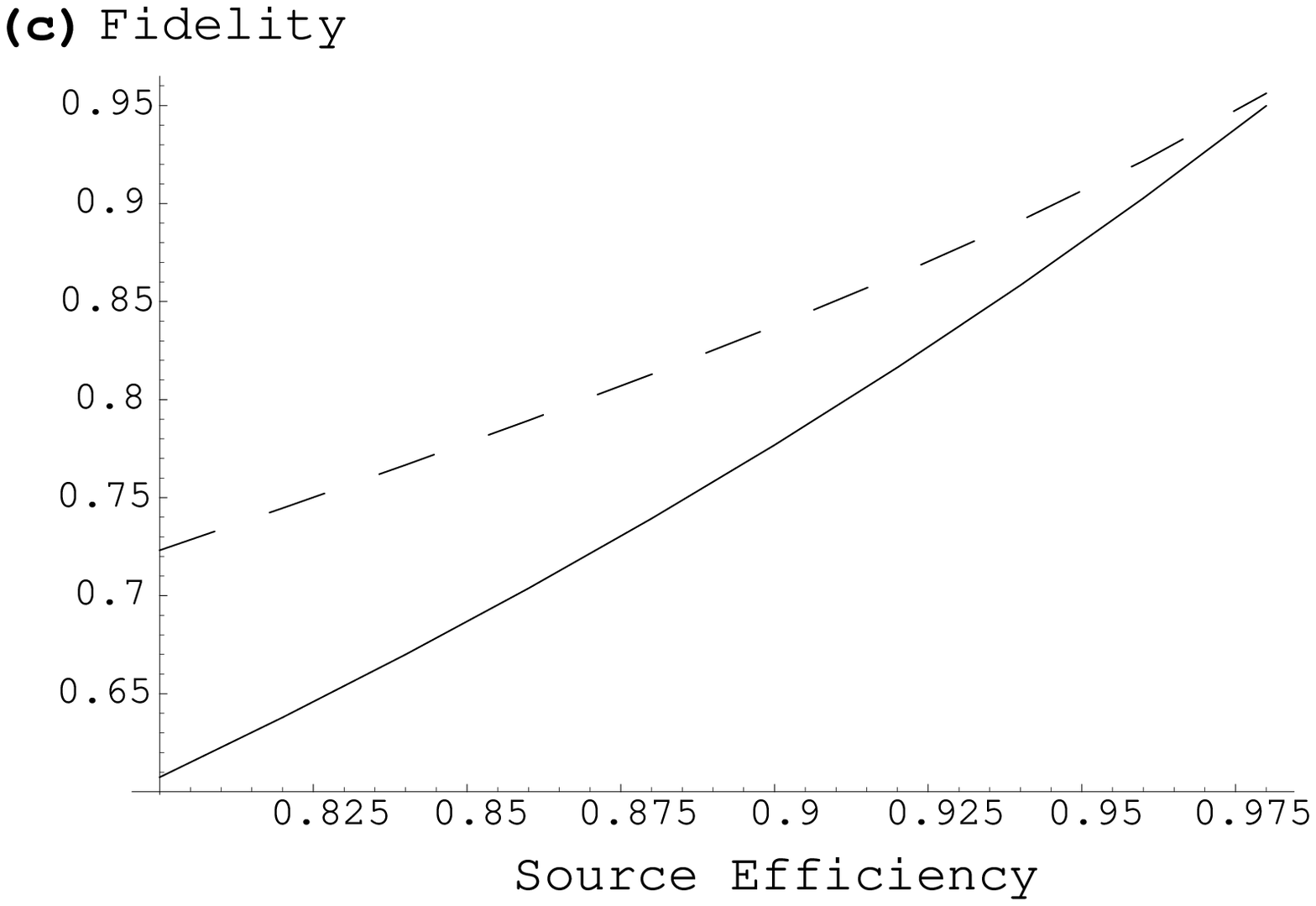}
  \includegraphics[width=0.4\textwidth]{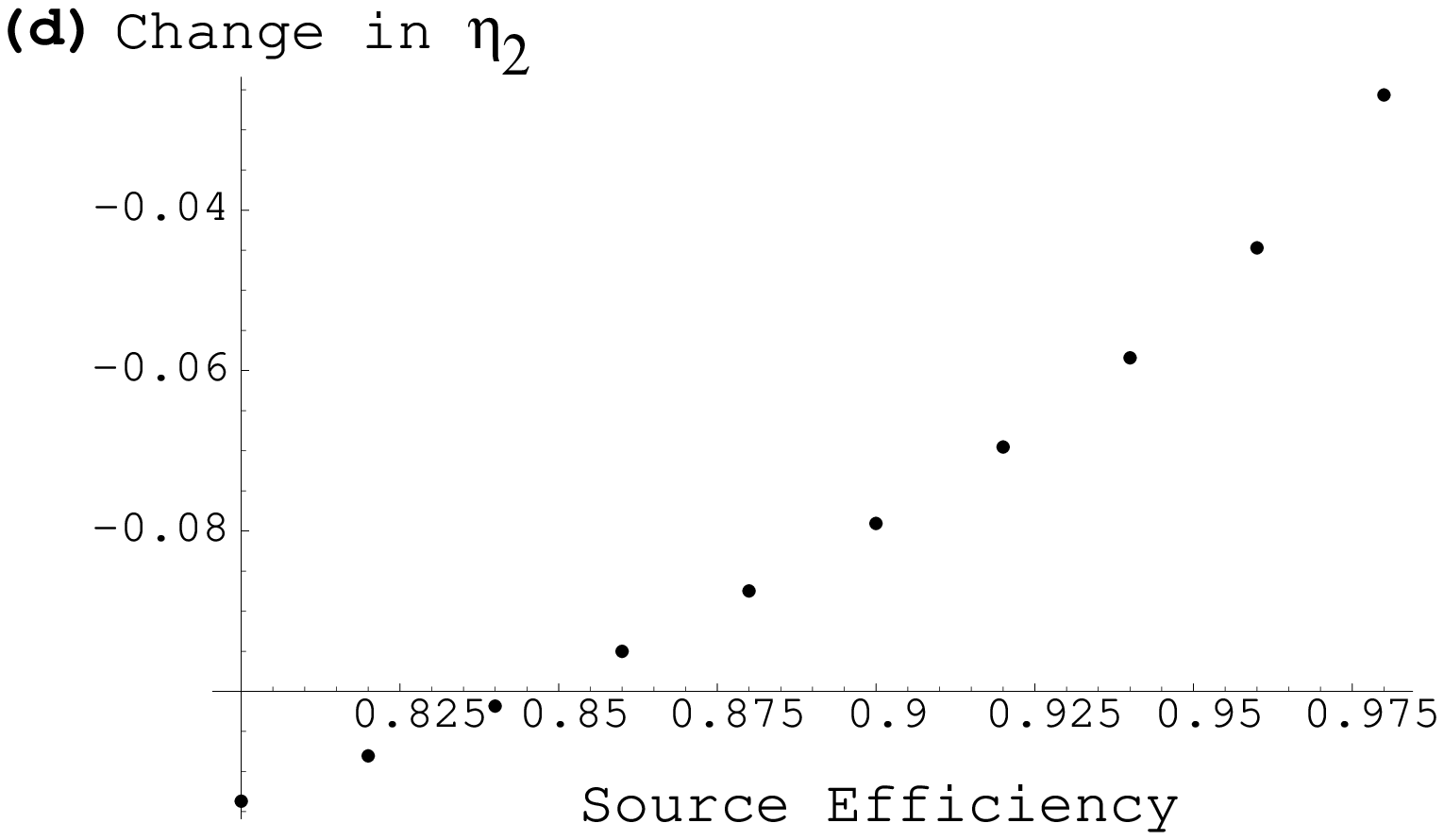}

  \includegraphics[width=0.4\textwidth]{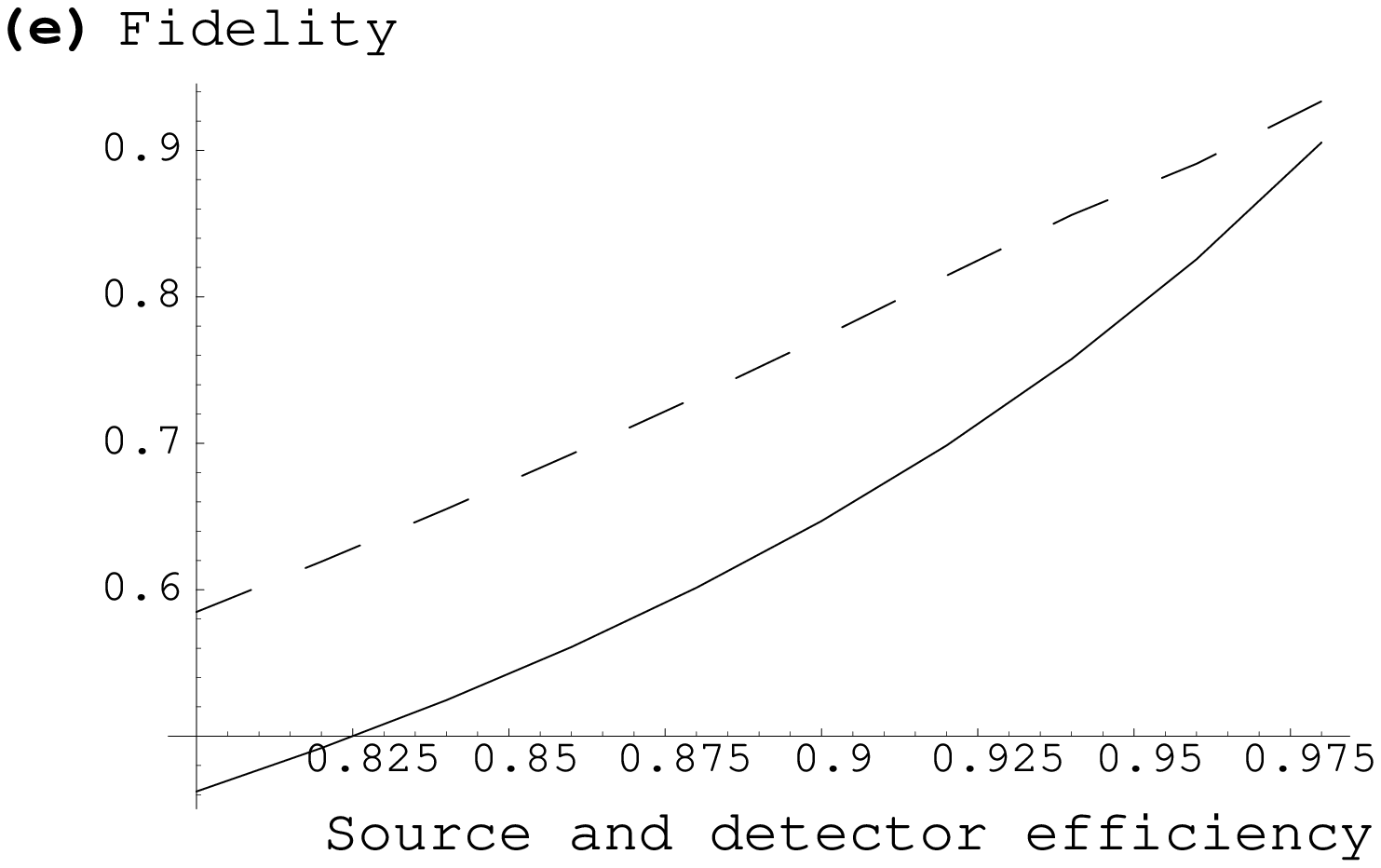}
  \includegraphics[width=0.4\textwidth]{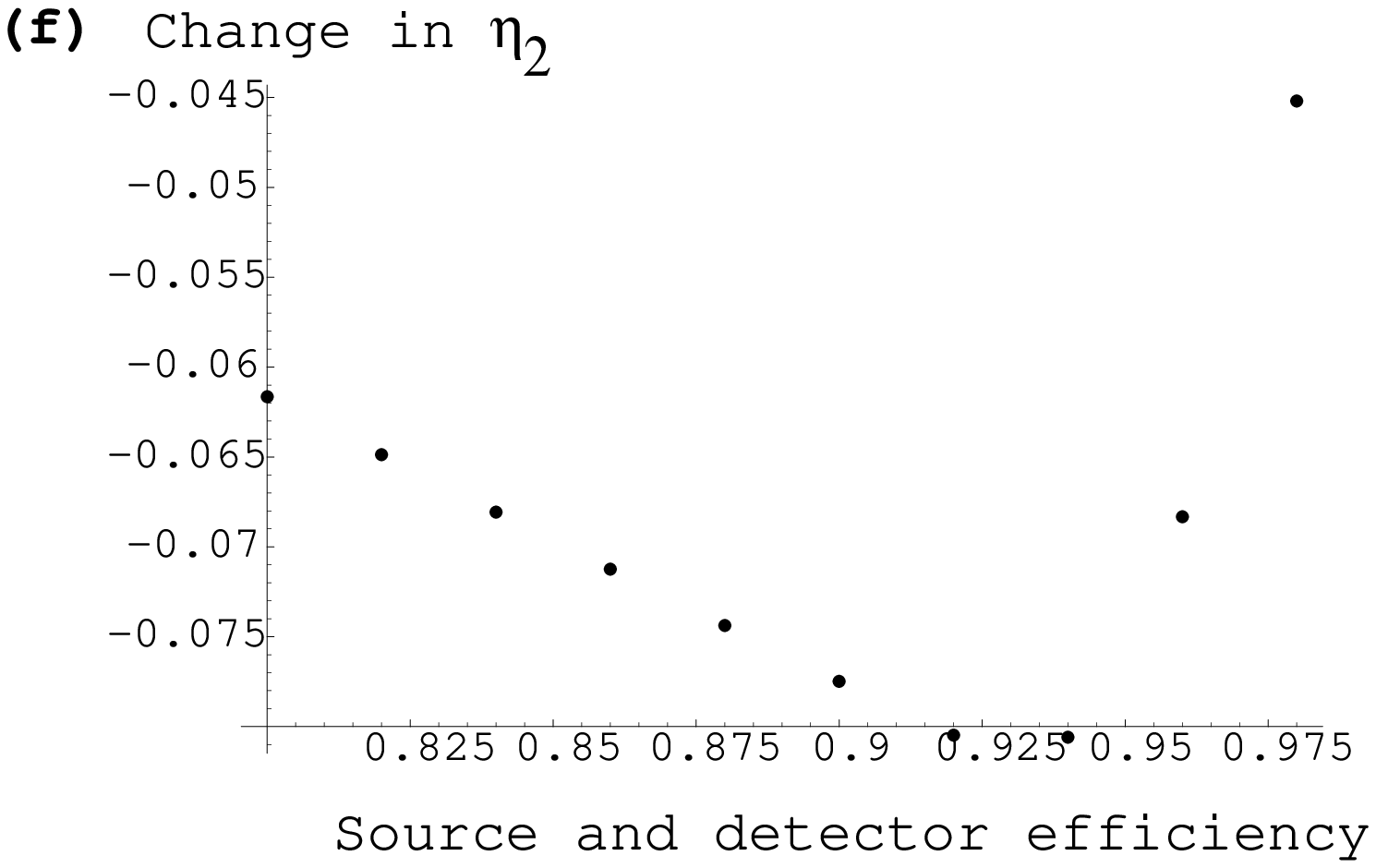}

  \caption{\label{imp1}The graphs in this figure show many cases in which the
  fidelity of the modified KLM gate can be improved.  The top row
  shows a simulation with detector efficiency considered only.  The
  graph on the left shows the fidelity of the gate without any
  adjustments with the solid line. The dashed line in this plot shows the
  fidelity
  reached when optimized against $\eta_{2}$ ($\eta_{1}=1$).  The amount
  that $\eta_{2}$ is changed by is shown in the plot on the right.
  This data is plotted with the detector efficiency along the abscissa.
  The other two sets of graphs show the same for source efficiency and
  finally detector and source efficiency both present but equal.}
\end{figure*}

Figure~\ref{imp1} shows the gate fidelity (optimized) with $\eta_{1} = 
1$ and $\eta_{2}$ at the optimum value.  The graphs on the left show 
the fidelity without any alterations to the beamsplitters (solid line 
and) the optimized fidelity (dashed line).  The plots on the right 
show what the $\eta_{2}$ value is for this optimized fidelity.  There 
are three cases shown in figure~\ref{imp1}.  The first is perfect 
source efficiency and variable detector efficiency.  The second is 
perfect detectors and variable source efficiency.  The last is 
variable source and detectors but both have equal efficiencies.

As an example of the small difference between using $\eta_{1}=1$ and 
varying it for non-unity source efficiency, the fidelity shown here 
for perfect detectors at 98\% source efficiency is 0.956.  When both 
$\eta_{1}$ and $\eta_{2}$ are varied a fidelity of 0.959 can be 
reached using $\eta_{1} = 0.7703$ and $\eta_{2}=0.1838$.  When a 
source efficiency of 0.8 is used, the fidelity reported here is 0.723 
and a slight improvement (in the fourth decimal place) can be achieved 
at the values $\eta_{1} = 0.9720$ and $\eta_{2}=0.1123$.  

\begin{figure*}
  \includegraphics[width=0.4\textwidth]{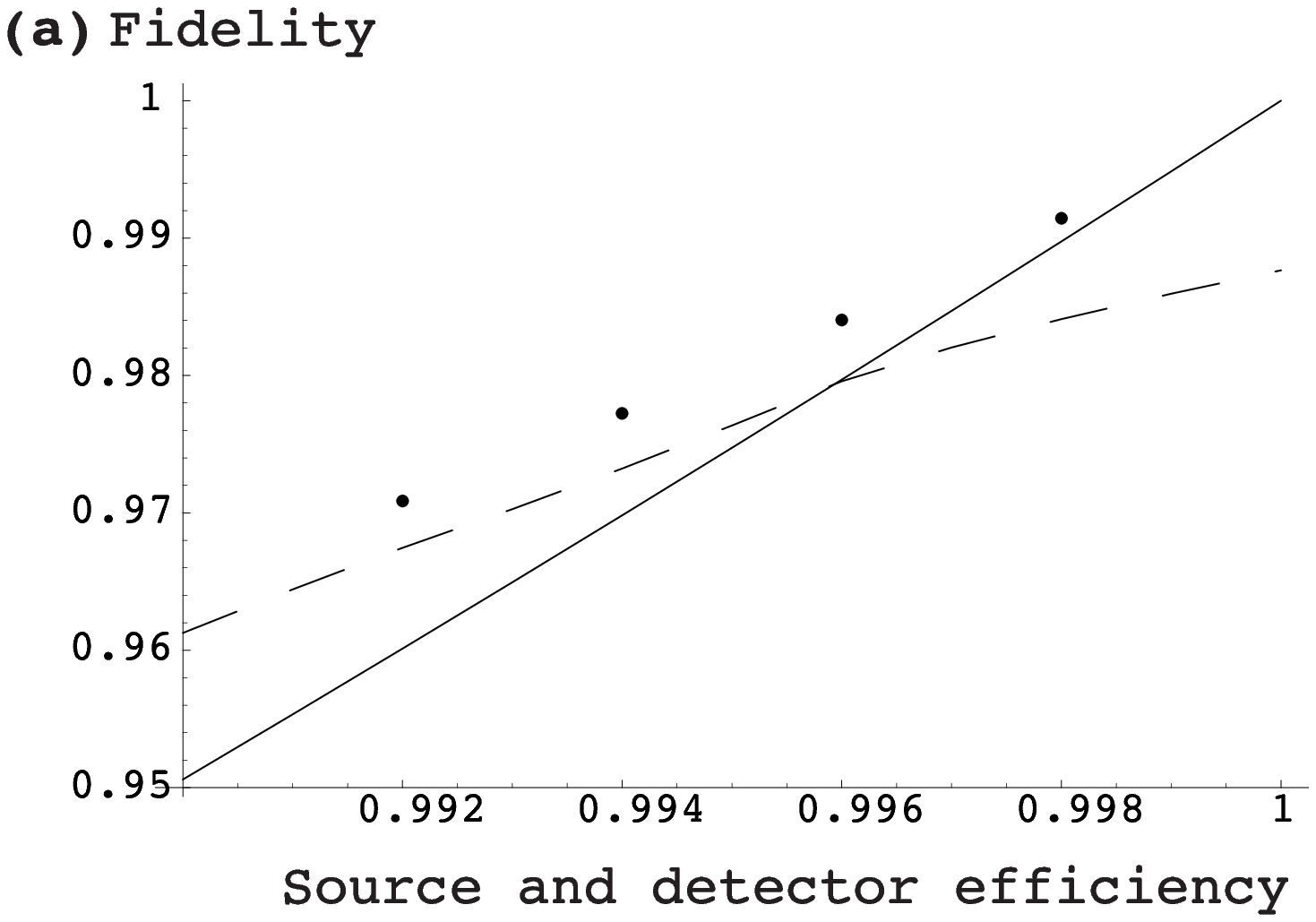}
  \includegraphics[width=0.4\textwidth]{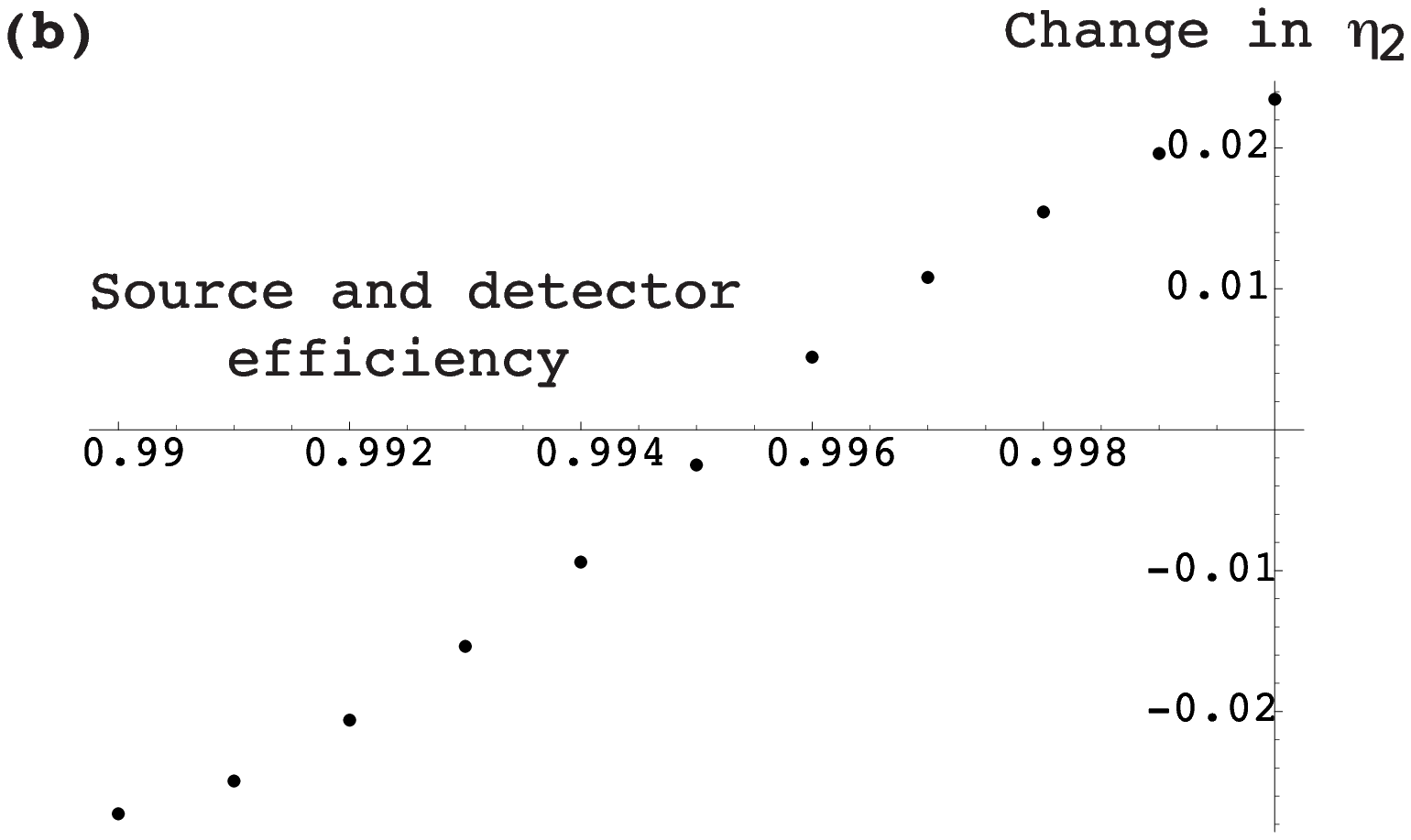}
  \caption{\label{imp2}These plots are of the same form as those in
  figure~\ref{imp1}.  They are a zoomed region of the detector and
  source losses considered (but equal) near unity efficiency.  This
  shows that setting $\eta_{1}=1$ can lead to an improvement until an
  efficiency of about 99.5\% is reached.  The dots shown on the left
  plot are the optimized fidelities when both $\eta_{1}$ and
  $\eta_{2}$ are varied.}
\end{figure*}

Figure~\ref{imp2} shows similar evidence that $\eta_{1}$ should be set 
to unity for all but the highest efficiencies.  The figure is a zoomed 
region of the plots from figure~\ref{imp1} where detector and 
source efficiencies are equal and higher than 0.99. 
Note from these figures that there is an improvement in
fidelity with $\eta_{1} = 1$ until efficiencies reach about 99.5\%.
Once again a slight improvement in these figures can be gained by
varying $\eta_{1}$ (which is possibly the origin of the slight
downwards bending of the improved fidelity curve).

\begin{figure}
  \includegraphics[width=0.4\textwidth]{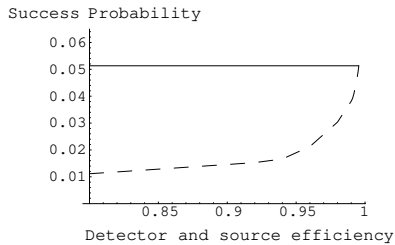}
  \caption{\label{success}This plot shows the probability of success
  of the KLM gate when at the optimized value for $\eta_{2}$ and setting
  $\eta_{1}=1$.  The case considered here is when detector and source
  losses are present and equal in magnitude.  The solid line shows the
  probability of success of the gate with the beamsplitter ratios set
  to the usual values and with no losses.}
\end{figure}

Changing the parameters of the gate will change the probability that 
the gate will function 
successfully as shown in figure~\ref{success} for 
the case where detector and source losses are equal.  The probability 
of the gate functioning does not drop below about $\frac{1}{5}$ the 
original value for detector and input efficiencies above about 0.8.

This technique of tuning gate parameters to counter the 
effects of ancilla inefficiency could also be applied to the Knill 
and PJF gates in 
some form.  However, it is not so clear how to proceed for these 
gates and it could be a computationally expensive task.  Since the KLM 
gate gave the most encouraging results in the default setup and its 
parameter space is relatively small, its optimization was pursued here.

\section{\label{conc}Conclusion}

Three LOQC C-sign gates have been compared using the minimum 
fidelity over all possible input states as the figure of merit.  The 
KLM gate appears to be the most resilient to photon loss in ancilla 
detection for efficiencies below 95\% and input loss for all
efficiencies.  The gate fidelity for the KLM gate can be improved by 
adjusting the beamsplitter ratios of the gate.  In all but the most 
efficient conditions (loss less than 0.5\%), it is best to remove the
first beamsplitter from each of the two NS gates that make up the 
C-sign gate and adjust the second until optimum fidelity is reached.  
This actually reduces the complexity of the gate considerably by 
removing two photon counters.  The improvement in minimum fidelity can 
be quite significant. Single photon production and detection 
efficiencies around 90\% are not unreasonable in the short term. Under 
such conditions the optimized KLM gate could be expected to give 
fidelities $\ge 0.8$ for all operations, assuming all other 
imperfections can be neglected.

\acknowledgments

We acknowledge useful discussions with G.~J.~Milburn and 
A.~Gilchrist. This work was supported by the Australian Research 
Council and ARDA.


\begin{thebibliography}{99}


\bibitem{kni00} E.~Knill, R.~Laflamme and G.~Milburn, Nature {\bf 409},
46, (2001).

\bibitem{glancy} S.~Glancy, J.~M.~LoSecco, H.~M.~Vasconcelos, and C.~E.~Tanner
\pra {\bf 65}, 062317 (2002)

\bibitem{ral00} T.~C.~Ralph, A.~G.~White, W.~J.~Munro and
G.~J.~Milburn, \pra, {\bf 65}, 012314 (2001).

\bibitem{kni01} E.~Knill, \pra {\bf 66}, 052306 (2002).

\bibitem{pit00} T.~B.~Pittman, B.~C.~Jacobs, and J.~D.~Franson,
\pra, {\bf 64}, 062311 (2001).

\bibitem{lun00} A.~P.~Lund and T.~C.~Ralph, \pra, {\bf 66}, 032307
(2002).

\bibitem{mil88} G.~J.~Milburn, \prl {\bf 62}, 2124 (1988).

\bibitem{koashi} M.~Koashi, T.~Yamamoto and N.~Imoto \pra {\bf 63}
030301 (2001).

\end{thebibliography}
\end{document}